\documentclass[traditabstract]{aa}
\usepackage{graphicx}                    
\usepackage{bm}
\usepackage{times}                    
\usepackage{color}                       
\usepackage{url}                         

\def\beg{\begin{eqnarray}}
\def\ende{\end{eqnarray}}
\def\lsim{\lower.4ex\hbox{$\;\buildrel <\over{\scriptstyle\sim}\;$}}
\def\gsim{\lower.4ex\hbox{$\;\buildrel >\over{\scriptstyle\sim}\;$}}

\newcommand{\Pm}{\mbox{Pm}}
\newcommand{\Rm}{\mbox{Rm}}
\newcommand{\Rey}{\mbox{Re}}
\renewcommand{\vec}[1]{\mbox{\boldmath $#1$}}
\def\curl{{\rm curl}} 
\def\Om{{\it \Omega}}
\def\R{{R\"udiger}}

\begin{document}



\title{Current helicity and electromotive force of  magnetoconvection influenced by    helical background fields}
\titlerunning{Current helicity and electromotive force}
%
\author{ G.~R\"udiger  \and M.~K\"uker}


%
  \institute{Leibniz-Institut f\"ur Astrophysik Potsdam, An der Sternwarte 16, D-14482 Potsdam, Germany,
                     email: gruediger@aip.de, mkueker@aip.de}

\date{Received; accepted}
 
\abstract{Motivated by the empirical  finding that the known   hemispheric rules for the current helicity at the solar surface are not  strict, the excitation of small-scale current helicity by the influence of  a large-scale helical magnetic background fields  on nonrotating magnetoconvection is demonstrated. It is shown within a quasilinear analytic theory of driven turbulence   and by nonlinear  simulations of magnetoconvection that the resulting small-scale current helicity has the same sign as the large-scale  current helicity  while the ratio of both pseudo-scalars is of the order of the magnetic Reynolds number of the convection. The same models do not provide  finite values of the small-scale kinetic helicity.  On the other hand,   a turbulence-induced electromotive force is produced including  the diamagnetic pumping term as well as the eddy diffusivity  but  no $\alpha$ effect.  It is thus argued  that the relations by  Pouquet \& Patterson (1978) and Keinigs (1983) for  the simultaneous existence of small-scale current helicity and $\alpha$ effect do not hold for the considered model  of nonrotating  magnetoconvection.  Calculations for various values of the magnetic Prandtl number demonstrate that for the considered  diffusivities the current helicity grows for growing magnetic Reynolds number which is not true for the velocity of the diamagnetic pumping -- in agreement with the results of the  quasilinear analytical approximation.
}


%

\maketitle
%
\section{Introduction} \label{Section1}
An increasing number of   
observations concerns  the small-scale  current helicity 
\beg
{\cal H}_{\rm curr}=\langle \vec{b}(\vec{x},t)\cdot  \curl\, \vec{b}(\vec{x},t)\rangle
\label{heli}
\ende
 at the solar surface, all showing that 
it is  negative (positive) in
the northern (southern) hemisphere  (Hale 1927; Seehafer 1990; Rust \& Kumar
1994;
 Abramenko et al. 1996: Bao \& Zhang
1998; Pevtsov 2001; Kleeorin et al.~2003; Zhang et al. 2010: Zhang 2012).  Here the notation $\vec b$ denotes the fluctuating parts of the total magnetic field in the form $\vec B= \bar{\vec{B}}+ \vec{b}$. It became clear that the scalar quantity (\ref{heli})  shows a strict equatorial
antisymmetry with signs which do not change from cycle to cycle.  Note that  for the linear $\alpha^2$ dynamo as well as the simple $\alpha\Om$ dynamos  the large-scale helicity    $\bar{\vec{B}}\cdot  \curl\, \bar{\vec{B}}$ similar to (\ref{heli})  also shows   equatorial antisymmetry  in accordance with  ${\rm sign}(\bar{\vec{B}}\cdot  \curl\, \bar{\vec{B}})={\rm sign}(\alpha)$ (Steenbeck \& Krause 1969). That  for flux transport $\alpha\Om$ dynamos the relation is much more complicated (positive correlation only during the cycle minima) indicates   that a direct relation $\alpha \propto \bar{\vec{B}}\cdot  \curl\, \bar{\vec{B}} $ not necessarily exists.
It is nevertheless important  that  all the mentioned scalars such as helicities and $\alpha$ effect are pseudoscalars which might be related to each other.

There are many theoretical studies where  both the current helicity and the $\alpha$ effect are derived as consequences of the existence of the pseudoscalar $\vec{g}\cdot \vec{\Om}$ in rotating convection zones with $\vec g$ as the (radial) direction of stratification. Yet it is shown in the present paper that the current helicity can also be produced even without rotation if the convection is influenced by a magnetic background field which  is  helical. Then the question  is whether the same constellation would also  lead to   a turbulence-induced  electromotive force via an $\alpha$ effect  as suggested by the relation 
\beg
\alpha=-\eta\frac{\langle \vec{b}(\vec{x},t)\cdot  \curl\, \vec{b}(\vec{x},t)\rangle
}{\bar{\vec{B}}^2}
\label{Kei}
\ende
of Keinigs (1983), It is based on the existence of homogeneous and stationary turbulence with finite kinetic helicity (Seehafer 1996) which conditions are obviously not fulfilled if one of the magnetic field components is inhomogeneous.   Pouquet \& Patterson (1978)   presented  the   relation
\beg
\alpha\propto\frac{1}{\mu_0\rho}\langle  { {\vec{b}}\cdot
{\curl}\,{\vec{b}}}\rangle -   \langle { {\vec{u}}\cdot
{\curl}\,{\vec{u}}}                 \rangle, 
\label{Pouq}
\ende
which also   means that  even turbulent fluids with vanishing kinematic helicity but finite small-scale current helicity should possess an $\alpha$ effect but with opposite sign as in (\ref{Kei}). 

Studies  by Yousef et al. (2003),  Blackman \& Subramanian (2013) and Bhat et al. (2014 ) concern   the  dissipation    of helical  and nonhelical large-scale background fields { under the influence of a nonhelical turbulent forcing. It is shown in these  papers that the  kinematic helicity is dominated by the magnetic helicity which   is  confirmed by our  quasilinear  calculations for forced turbulence and the nonlinear simulations of magnetoconvection 
 under the influence of a large-scale helical field.  In contradiction to the relation (\ref{Pouq})  both   the quasilinear approximation {\em and}  nonlinear simulations provide reasonable values for the current helicity but they do not lead to a finite $\alpha$ effect.}

 \section{The current helicity}\label{Section2} 
In order to find the current helicity  due to the interaction of a prescribed  stochastic velocity  $\vec{u}(\vec{x},t)$ and 
a large scale magnetic field $\bar{ \vec{B}}(\vec{x})$ it is sufficient  to solve the  induction equation
\begin{equation}
\frac{\partial\vec{B}}{\partial t}  = {\rm curl}\,(\vec{u}\!\times\!\vec{B})+
\eta\  \Delta\vec{B} ,
\label{1}
\end{equation} 
where the fluctuating magnetic field may be written as $\vec{B}=\bar{\vec{B}}+ \vec{b}$.
The influence of field gradients on the mean current helicity (\ref{heli}) 
 at linear order is governed by 
\beg
\frac{\partial {b_i}}{\partial t}- \eta \, \Delta {b_i}
= x_p B_{jp} u_{i,j} - u_j B_{ij},
\label{partb}
\ende
where the prescribed  inhomogeneous mean magnetic field $\bar{\vec B}$ has been introduced in the form
$
\bar B_j= B_{jp} x_p
$
with the notation  $\bar{B}_{j,p}=B_{jp}$ and with $B_{jj}=0$. It follows $\bar{\vec{B}}\cdot {\curl}\, \bar{\vec{B}}=
\epsilon_{ilk} B_{ij} B_{kl} x_j$ which is here considered to be the only pseudo-scalar in the
system.

To solve the   equation (\ref{partb})  the use of the inhomogeneous Fourier modes
\begin{eqnarray}
u_i (\vec{x},t)&=& \int\!\!\!\!\int \hat u_i (\vec{k},\omega)
 e^{{\rm i}({\bm k}{\bm x}-\omega t)} {\rm d}\vec{k} \, {\rm d}\omega, 
\nonumber\\
b_i (\vec{x},t)& =& \int\!\!\!\!\int (\hat b_i (\vec{k},\omega)
 + x_l \hat b_{il} (\vec{k},\omega)) e^{{\rm i}({\bm k}{\bm x} -\omega t)}
 {\rm d} \vec{k} \, {\rm d}\omega
\label{uibi}
\end{eqnarray}
is suggestive. It is a  standard procedure which  yields 
\beg
\hat b_i = -\frac{B_{ij} + \frac{2\eta k_l k_m B_{lm} \delta_{ij}}
{-{\rm i}\omega + \eta k^2}}{-{\rm i} \omega + \eta k^2} \hat u_j,
 \quad\quad \quad \hat b_{il}= \frac{{\rm i}k_j B_{jl}}{-{\rm i} \omega
 + \eta k^2} \hat u_i
\label{bihatbi}
\ende
 (\R\ 1975)  from which  the small-scale current helicity  
\begin{eqnarray}
{\cal H}_{\rm curr}&= &\int\!\!\!\!\int 
\langle \hat{\vec{b}}(\vec{k},\omega)\cdot  \curl\, \hat{\vec{b}}(\vec{k'},\omega')\rangle
 e^{{\rm i}(({\bm k}+{\bm k'}){\bm x}-(\omega+\omega')t)} \nonumber\\
&&\ \ \ \ \ \ \ \ \ \ \ \ \ \ \ \ \ \ \ \ \ \ \ \ \ \ \ \ {\rm d}\vec{k} {\rm d}\vec{k'}\, {\rm d}\omega  {\rm d}\omega' 
\label{hel1}
\end{eqnarray}
can easily be formed. A homogeneous and isotropic turbulence field with the  spectral tensor 
\beg 
\hat Q^{(0)}_{ij}(\vec{k}, \omega)=\frac{E(k,\omega)}{16\pi k^2}
 \left(\delta_{ij}-\frac{k_i 
k_j}{k^2}\right) 
\label{qu} 
\ende 
may be postulated where the positive  spectrum $E$ gives the intensity of the  fluctuations. 
After manipulations  one derives  from   (\ref{hel1}) and (\ref{qu}) the final expression
\begin{equation}
{\cal H}_{\rm curr}= \frac{1}{6} \int\limits_0^\infty
 \int\limits_{-\infty}^\infty 
\frac{k^2 E(k,\omega)}{\omega^2+\eta^2 k^4}
{\rm d}k {\rm d}\omega \ \ \bar{\vec{B}}\cdot{\rm curl}\, \bar{\vec{B}}.
\label{calh5}
\end{equation}
The integral in
(\ref{calh5}) does only  exist for the high-conductivity limit $\eta \to 0$ after multiplication with $\eta$ so that  
\beg
{\cal H}_{\rm curr}\simeq \frac{ \Rm}{3}\  \bar{\vec{B}}\cdot
{\curl}\,\bar{\vec{B}}
\label{HH}
\ende
scales with the magnetic Reynolds number  $\Rm=u_{\rm rms}\ell_{\rm corr}/\eta$.

   Equations (\ref{Pouq}) and (\ref{HH})   lead to  
\beg
\alpha\propto-  \langle { {\vec{u}}\cdot
{\curl}\,{\vec{u}}}\rangle+\frac{\Rm}{3\mu_0\rho}\  { {\bar{\vec{B}}}\cdot
{\curl}\,{ \bar{\vec{B}}}}                    
\label{BS}
\ende
for the $\alpha$ effect for weak magnetic fields ({ see Blackman \& Brandenburg 2002},  Brandenburg \& Subramanian 2005).

With  our approximations ({ quasilinear equations, forced turbulence, no rotation}) the ratio of the  two  helicities  corresponds to the ratio of the both magnetic  energies.
Equation  (\ref{HH}) requires  the same sign for  the small-scale and the large-scale current  helicities.  With (\ref{Kei}) one finds ${\rm sign}({\alpha})=-{\rm sign}({ \bar{\vec{B}}\cdot
{\curl}\,\bar{\vec{B}}})$ which  contradicts the above mentioned observations that in the simple $\alpha^2$ dynamos 
$ {\rm sign}({\bar{\vec{B}}\cdot
{\curl}\,\bar{\vec{B}}})= {\rm sign}({\alpha})$. The only solution of this dilemma is that nonrotating  convection under the influence of helical large-scale fields may produce small-scale current helicity but it does not generate   an $\alpha$ effect. 

{ If calculated under  the same   analytical assumptions  as the current helicity,  the kinetic helicity $\langle  {\vec{u}}\cdot
{\curl}\,{\vec{u}}   \rangle$  identically   vanishes.  Blackman \& Subramanian (2013) demonstrated  that during the turbulence-induced decay of helical large-scale fields the kinetic helicity remains unchanged. 
In the quasilinear approximation its part  due to the background  fields is identically  zero.  MHD turbulence  under the influence of a helical large-scale  field is thus suspected to generate current helicity but no  kinetic helicity. The question is whether on this  basis and after  the relations (\ref{Kei}) and/or (\ref{Pouq}) also nonlinear  simulations of convection in a stratified plasma  provide current helicity but no kinetic helicity  and no  $\alpha$ effect.}

In a previous paper we simulated  magnetoconvection under the influence of a uniform vertical  magnetic  field in order to produce finite values of cross helicity (\R\ et al. 2012, from now on referred to  as paper I) .  In the present paper    the {\em helical} magnetic background field 
\beg
B_x= {\bar B}_x \frac{z}{H} \ \ \ \ \ \ \ \ \ \ \ \ \ \ \ \ \ \ \ \ \ \ \  \ B_y=  \bar B_y
\label{BB}
\ende
 with uniform values of ${\bar B}_x$ and $\bar B_y$ has been applied to probe the possible production of both small-scale kinetic and current helicity. In the computational layer the vertical coordinate $z$ runs from 0 to $2H$ while the convection zone is located in   [H,2H]. The value of the  electric-current is $J_y={\bar B}_x/(\mu_0 H)$. It is    $\bar{\vec{B}}\cdot {\curl}\, \bar{\vec{B}}\propto  \bar B_y {\bar B}_x$ by definition.  

\begin{figure}[h]
\center
\includegraphics[width=4.4cm,height=4cm]{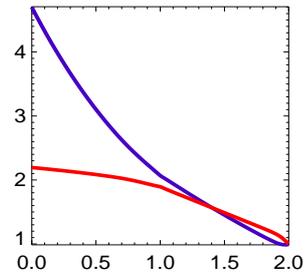}
\caption{Snapshots of temperature  (red lines) and density (blue lines) during the run for  ${\bar B}_x =10$, $\bar B_y=10$. The surface values are fixed to unity.}
\label{sound}
\end{figure}

The compressible MHD equations have been solved with the {\sc NIRVANA} code (Ziegler 2004) in a Cartesian box with gravitation along the negative $z$-axis. The box is periodic  in the horizontal directions and  all mean-field  quantities are averaged over the horizontal plane.  For  the Prandtl number and the magnetic Prandtl number a common value is fixed to 0.1. As in  Paper I the velocities are used in units of $c_{\rm ac}/100$ where $c_{\rm ac}$ is the speed of sound at the top of the convection box (see Fig. \ref{sound}). Correspondingly,  the magnetic fields are given in units of $\sqrt{\mu_0\rho}c_{\rm ac}/100$ with $\mu_0=4\pi$. We again assume an ideal, fully ionized gas heated from below and kept at a fixed temperature at the top of the simulation box. As is paper I the dimensionless Rayleigh number is fixed to $10^7$. Periodic boundary conditions apply at the horizontal boundaries.    The upper and lower boundary are   perfect conductors so that there  for the total magnetic fields 
\beg
\frac{{\rm d}B_x}{{\rm d} z}=\frac{{\rm d} B_y}{{\rm d}z} =B_z=0,
\label{BC}
\ende 
similar to the stress-free  conditions of the velocity, ${\rm d}u_x/{\rm d} z={\rm d} u_y/{\rm d}z =u_z=0$. At the inner boundary the heat-flux is fixed. One finds in paper I that for very weak magnetic fields the resulting rms-value of the convection velocity in the used units is about 8 (see also the left panel of Fig. \ref{fig1} below) so that with the dimensionlwss $\eta=0.06$ the {\em global} magnetic Reynolds slightly exceeds  100 which is  larger by definition than the   local magnetic Reynolds number  $\Rm=u_{\rm rms}\ell_{\rm corr}/\eta$   introduced above. 

We start to compute   the intensities    of the magnetoconvection   $u^2_{\rm rms}=\langle \vec{u}^2\rangle $ and $b^2_{\rm rms}=\langle \vec{b}^2\rangle $  (Fig. \ref{fig1}). The velocity at the top of the convective box results as   $\lsim 10$\% of the surface value of the sound speed  which for the Sun is about 10 km/s. The given  time averages and vertical averages in $z$=[1,2] of all snapshots  are taken as the characteristic values of the quantities. The differences of the velocity dispersion to those given in paper I, where much weaker vertical fields have been applied, are very small. The magnetic quenching of the velocity fields does exist but it is  weak. However, the ratio ${\rm q}=\langle \vec{b}^2\rangle/\vec{\bar B}^2 $ is strongly affected by the mean magnetic field. While  its value is about 30 for the vertical field ${\bar B}_z=1$ in paper I the helical field 
${\bar B}_x =\bar B_y=10$ only allows $\rm q$ values of order unity. Note also 
that  $u^2_{\rm rms}\gsim b^2_{\rm rms}/4\pi\rho$, i.e. the convection is {\em not} magnetic-dominated. Contrary to that, we shall demonstrate that nevertheless the magnetic helicity strongly dominates  the kinetic helicity.
\begin{figure}[h]
\hbox{
\includegraphics[width=4.4cm,height=4cm]{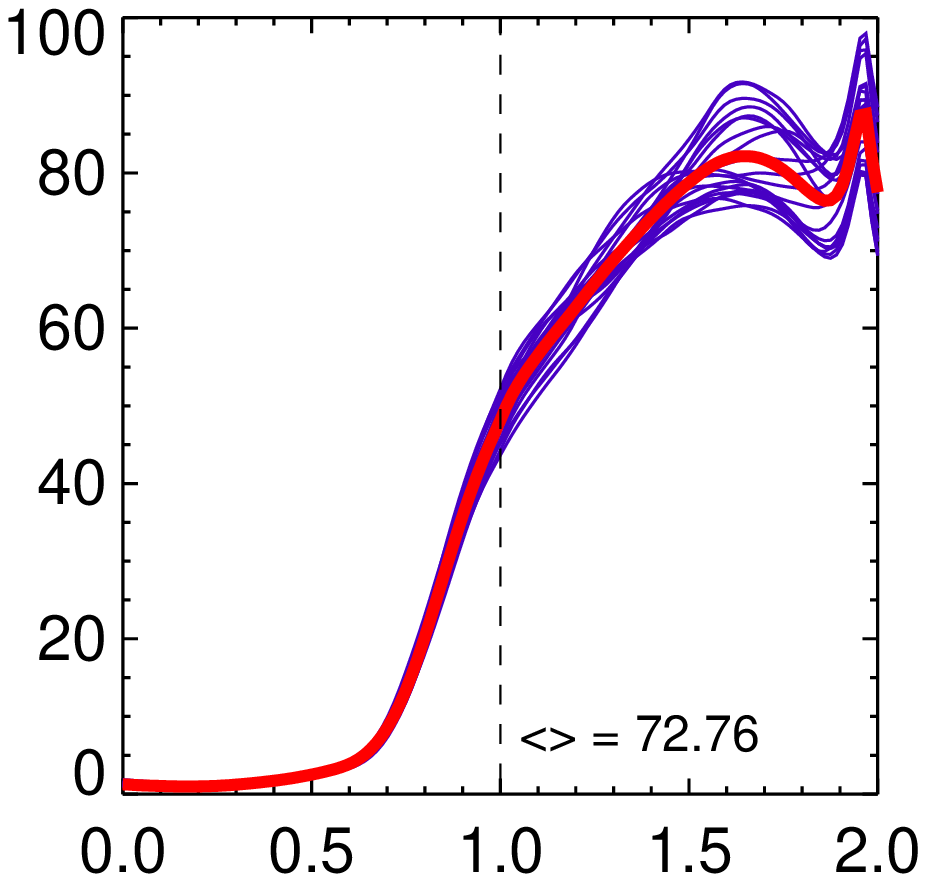}
\includegraphics[width=4.4cm,height=4cm]{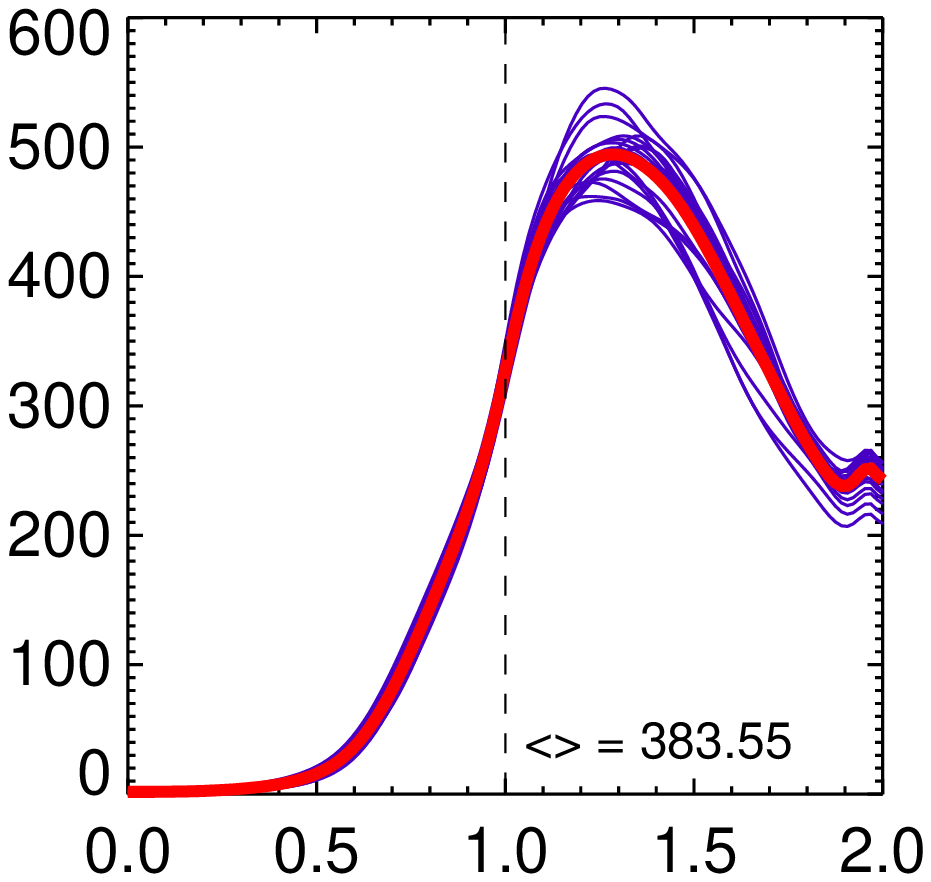}}
\caption{The turbulence-intensities    $u^2_{\rm rms}$ (left) and $b^2_{\rm rms}$ (right) for ${\bar B}_x =10$, $\bar B_y=10$.  The  numbers in the graphs  are the volume averages of the coefficients over the whole convective domain. The averages over all blue-line snapshots  here and in all the below plots are given  by   solid red lines.}
\label{fig1}
\end{figure} 

\begin{figure}[h]
\hbox{
\includegraphics[width=4.4cm,height=4cm]{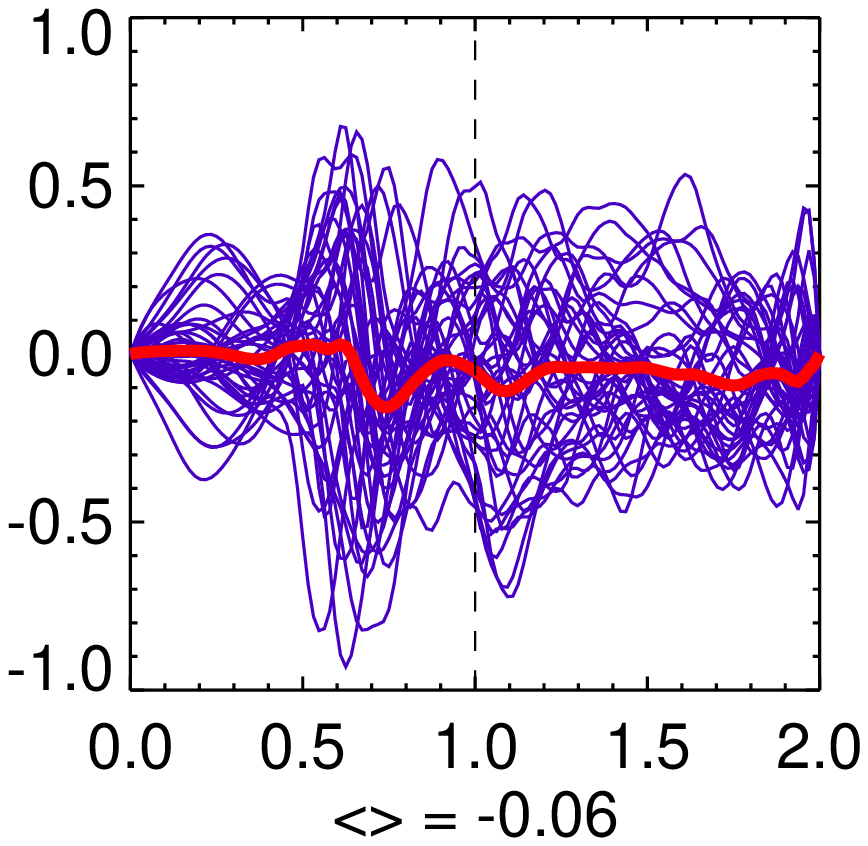}
\includegraphics[width=4.4cm,height=4cm]{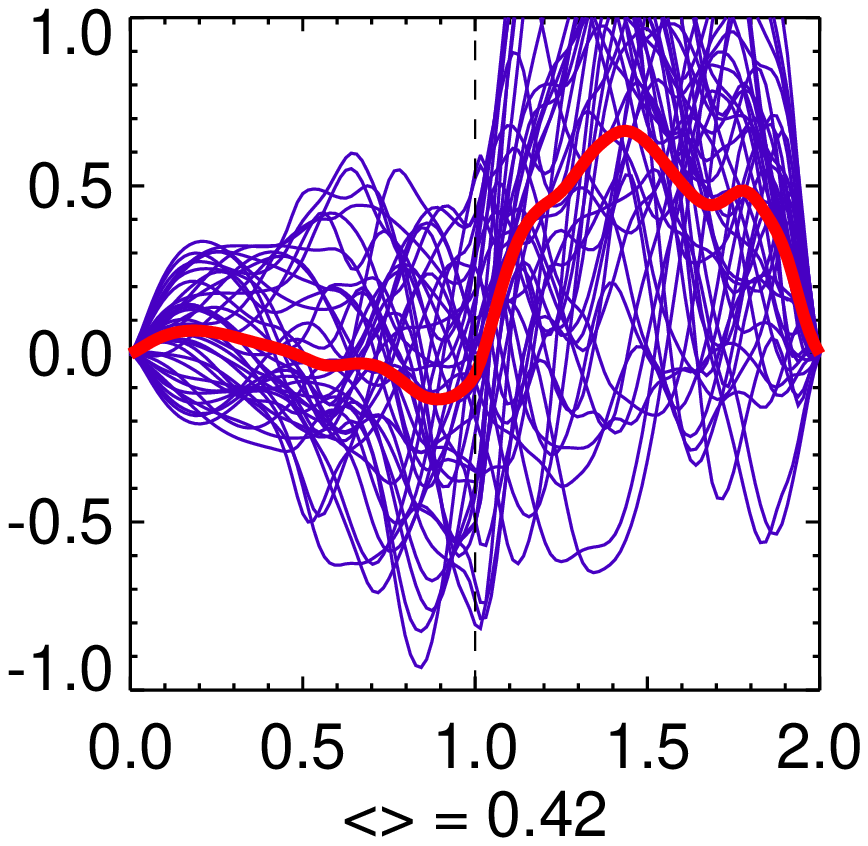}
}
\caption{The same as in Fig. \ref{fig1} but for the correlation coefficients $c_{\rm kin}$ (left)  and $c_{\rm curr}$ (right).   The  numbers below the graphs  are the volume averages of the coefficients over the whole convective domain.The average of all snapshots in both cases is below unity.}
\label{fig1c}
\end{figure}

For the same model  the correlation coefficients 
\beg
c_{\rm kin} = \frac{H \langle  {\vec{u}}\cdot
{\curl}\,{\vec{u}} \rangle }{u^2_{\rm rms}}, \ \ \ \ \ \ \ \ \  \ \ \ \ \ \ \ \ \  \ c_{\rm curr} = \frac{ H \langle  {\vec{b}}\cdot
{\curl}\,{\vec{b}}   \rangle }{b^2_{\rm rms}}
\label{BB}
\ende
for   the kinetic helicity and the current helicity have been computed (Fig. \ref{fig1c}).  The numerical values strongly differ. While the fluctuations of $\vec b$  and $\curl\, \vec b$ are well correlated this is not true for the  flow field. 

A   systematic behavior of the $z$-profile does also not exist.    In accordance with the analytic quasilinear calculations magnetoconvection subject to helical background fields does not generate small-scale kinetic helicity but it generates small-scale current helicity. 
\begin{figure}[h]
\hbox{
\includegraphics[width=4.4cm,height=4cm]{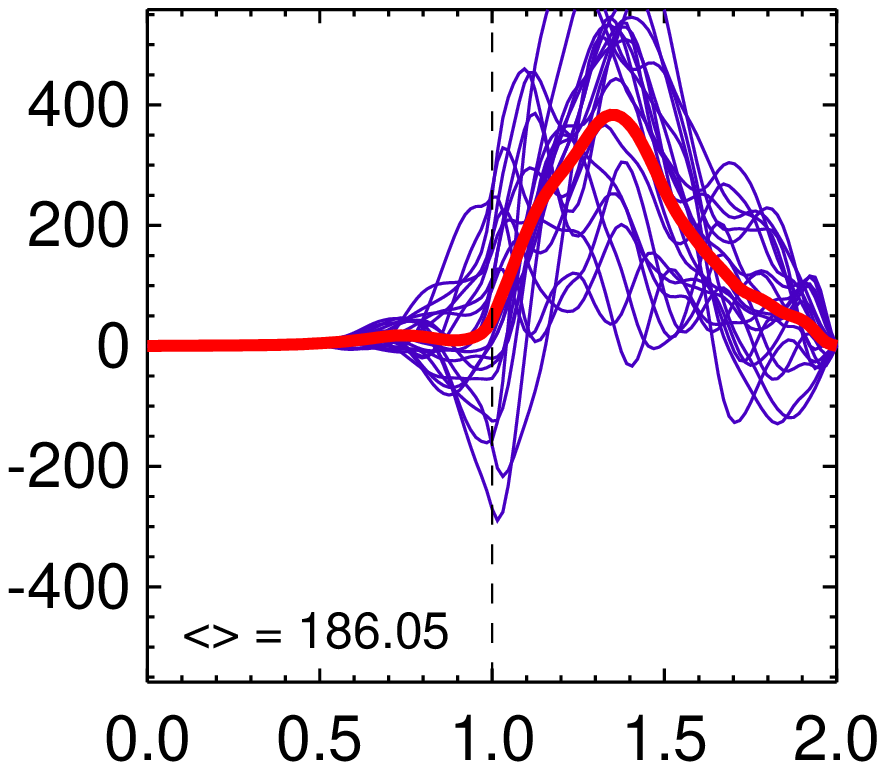}
\includegraphics[width=4.4cm,height=4cm]{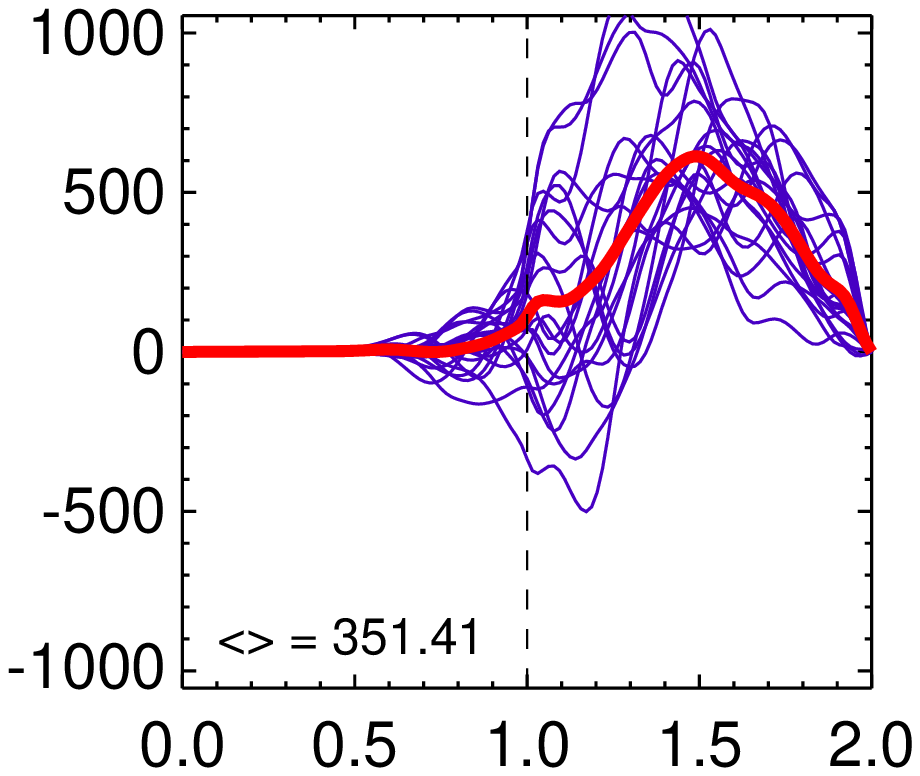}
}
\caption{The small-scale current helicity $\langle \vec{b}\cdot  \curl\, \vec{b}\rangle$  as due to the large-scale current helicity. Left panel: ${\bar B}_x =10$, $\bar B_y=10$, right panel: ${\bar B}_x=10$, $\bar B_y=20$. The numbers in the plots mark the averages of all given snapshots. The stratified layer  is convectively unstable at the right-hand side of the vertical dashed line where $z>1$.  }
\label{fig1b}
\end{figure}

 { The behavior of the magnetoconvection   complies  with  the above analytical result of the incapability of large-scale helical fields to produce small-scale kinetic helicity and they are in correspondence with  the numerical result of Bhat et al. (2014) who found the helicities as magnetic-dominated in their simulations of decaying helical background fields under the influence of nonhelically driven turbulence.}

After (\ref{HH})  one expects for the  current helicity numerical  values  exceeding     ${\bar B}_x \bar B_y$    which  indeed is    confirmed by the simulations (Fig. \ref{fig1b}).  The average  number of   the right plot is just  twice the number  in the left plot  demonstrating   the linear run  of the correlations with the  the large-scale helicity. The relation (\ref{HH}) which linearly connects the small-scale current helicity with the large-scale current helicity can thus be considered as realized  by the numerical simulations. The numerical value between the two helicities and its relation to the magnetic Reynolds number will be discussed below in the section \ref{Prandtl}.
\section{The electromotive force}\label{Section2} 
The same magnetoconvection model is used to calculate the turbulence-induced electromotive force. As electric-currents exist in the convection zone it certainly will contain tensorial components of magnetic diffusion. The question is whether also an electromotive force along the magnetic field occurs which must be interpreted as an $\alpha$ effect.  Surprisingly, the answer will be that no $\alpha$ effect is generated by the convection influenced by the helical magnetic background field (\ref{BB}). As simultaneously  current helicity without kinetic helicity exists, Eq. (\ref{Pouq}) cannot be correct  for all possible turbulence fields.

The electromotive force 
$
\vec{\cal E}\!=\langle  {\vec u} \times {\vec b}  \rangle
$
in  turbulent plasma is a polar vector. If one wants to express it linearly by the axial mean magnetic field vector  $\bar{\vec{B}}$ then only the formulation 
\begin{equation} 
\vec{\cal E}\!=- \eta_{\rm T}  {\rm curl} \bar{\vec{B}} - \vec{\gamma} \times  {\rm curl}\bar{\vec{B}} + ....
\label{E2} 
\end{equation} 
is possible if   pseudoscalars such as $\vec{g}\cdot \vec{\Om}$ or $\bar{\vec{B}} \cdot {\rm curl}\bar{\vec{B}}$ are not available. Without global rotation an $\alpha$ effect in the formulation $\vec{\cal E}= \alpha\ \bar{\vec{B}}\dots$ does not occur in the  linear relation  (\ref{E2}). The  parameter $\eta_{\rm T}$  describes the turbulence-induced eddy diffusivity in the fluid while the vector  $\vec{ \gamma}$ gives a velocity which transports (`pumps') the large-scale magnetic field downward for positive $\gamma_z$. 
If, however, terms nonlinear in $\bar{\vec{B}}$ are included in the heuristic expression (\ref{E2}) then the extra terms
\begin{equation} 
\vec{\cal E}\!=....+{\hat\eta}\ \ 
\bigg(\!\bar{\vec{B}}^2 \, {\rm curl}\, \bar{\vec{B}}
 + \frac{1}{3} \bigg[\nabla \bar{\vec{B}}^2 + (\bar{\vec{B}}\cdot\nabla)
 \bar{\vec{B}}\bigg]\times \bar{\vec{B}}\!\bigg) 
\label{E3} 
\end{equation} 
 occur within the second-order-correlation approximation (Kitchatinov  et al.\ 1994).  The first term of the RHS 
represents the known  magnetic influence on the  eddy diffusivity 
while the second one  forms a  magnetic-induced pumping effect. There is no term   $(\bar{\vec{B}} \cdot {\rm curl}
\bar{\vec{B}})\bar{\vec{B}}$ representing an $\alpha$ effect with the
large-scale current helicity $\bar{\vec{B}} \cdot {\rm curl} \bar{\vec{B}}$ of the
 background field as (pseudoscalar) coefficient.  For
 nonrotating and  non-helically driven turbulence the large-scale current helicity does not create an 
 $\alpha$~effect in this approximation. Indeed, we do not find an $\alpha$ effect in the  numerical simulations below. 
 There is instead the -- sometimes  ignored -- magnetic-induced new pumping term which for positive $\hat \eta$ advects  the magnetic field 
 towards   the maximum magnetic field.   This effect should appear in all  numerical models with  helical large-scale field and nonhelically  driven  flow (Yousef et al. 2003;  Bhat et al.  2014).

Equations (\ref{E2}) and (\ref{E3}) yield
 \begin{equation} 
\vec{\cal E}\!=-( \eta_{\rm T}-\hat\eta \frac{{\bar{\vec B}}^2}{\mu_0\rho})  {\rm curl} \bar{\vec{B}} - (\vec{\gamma}- \frac{\hat\eta}{3} \frac{\nabla {\bar{\vec  B}}^2}{\mu_0\rho} )\times  \bar{\vec{B}}.
\label{E4} 
\end{equation} 
 The nonlinear terms magnetically affect   both the eddy diffusivity and the advection velocity which 
  are basically   reduced if the factor  
\begin{equation} 
{\hat \eta}\!=\! \frac{1}{5} \int\limits_0^\infty \!\!
\int\limits_{-\infty}^\infty \frac{\eta k^4(\nu\eta \, k^4
 - \omega^2)E(k,\omega)\ {\rm d}k \, {\rm d}\omega}{\left(\omega^2+\eta^2 k^4\right)^2
\left(\omega^2+\nu^2 k^4\right)} 
\label{E5} 
\end{equation} 
is positive. 
The $\hat\eta$ possesses the dimension of a time and can be estimated as $\hat\eta\simeq \Rm^{2} \tau_{\rm corr}$ in the high-conductivity limit. One easily finds  $\hat\eta>0$ for important cases. For $\nu=\eta$ 
the coefficient 
$\hat \eta$ is always positive for all spectral functions which do not grow with growing frequency. Moreover,  for all $\Pm$ the full expression (\ref{E5}) is positive  for  $E\propto \delta(\omega)$ representing long correlation times as well as for $E \simeq {\rm  const}$, i.e.very short correlation times (`white noise').  If compared with the magnetic-diffusion time $\mu_0\sigma \ell^2_{\rm corr}$ the delta-like spectral line represents the low-conductivity limit while the white-noise spectrum represents the  high-conductivity limit.  In both limits the $\hat\eta$ is positive and, therefore,  both eddy diffusivity and diamagnetic pumping are magnetically quenched rather than amplified (`anti-quenched').

\subsection{Numerical simulations}
The two  horizontal  components of the  electromotive force  are calculated for two different cases. In the first simulations  the applied magnetic fields are assumed as free of 
helicity, i.e. $\bar{\vec{B}} \cdot {\curl} \bar{\vec{B}}=0$. 
The other possibility works with  $\bar{\vec{B}} \cdot {\curl} \bar{\vec{B}}\neq 0$. If the results of both setups are identical then the global helicity does not contribute to 
the electromotive force as is suggested by Eq. (\ref{E3}).

\begin{figure}[hbt]
\begin{center} 
\hbox{
\includegraphics[width=4.5cm,height=6cm]{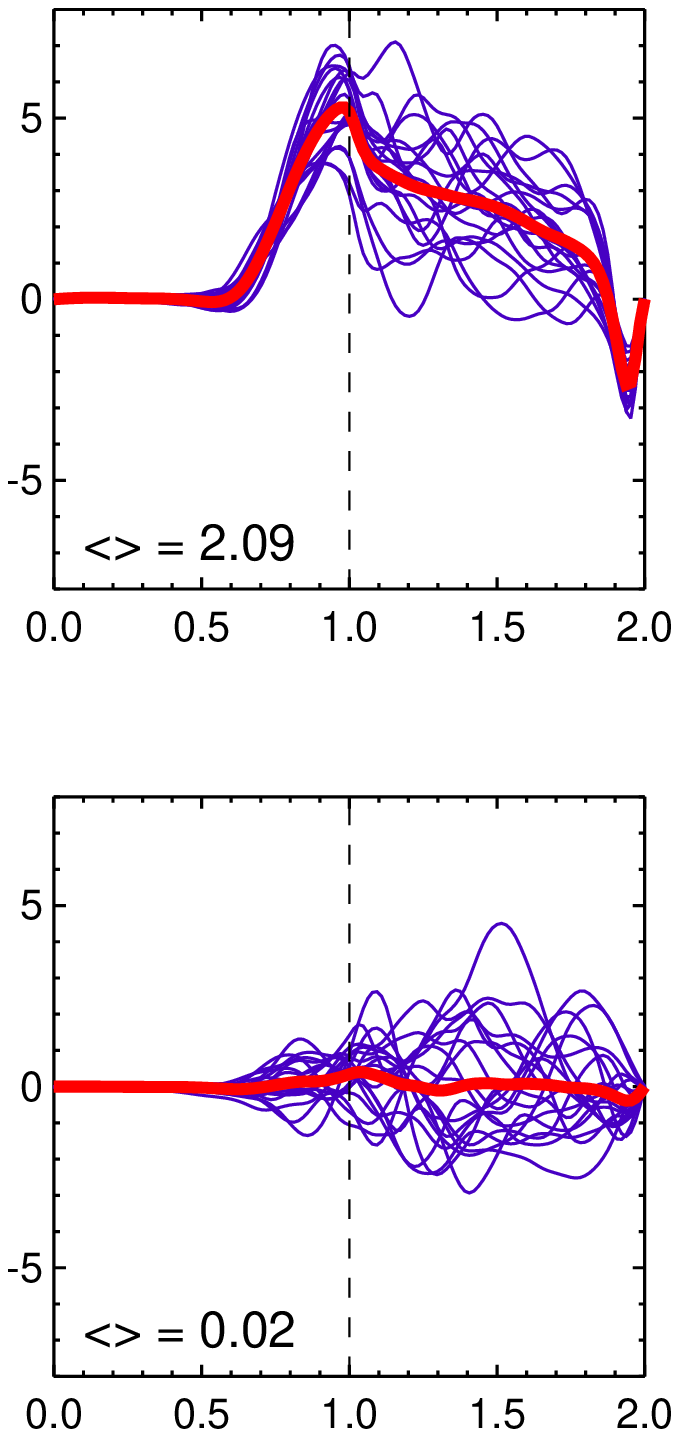}
\includegraphics[width=4.5cm,height=6cm]{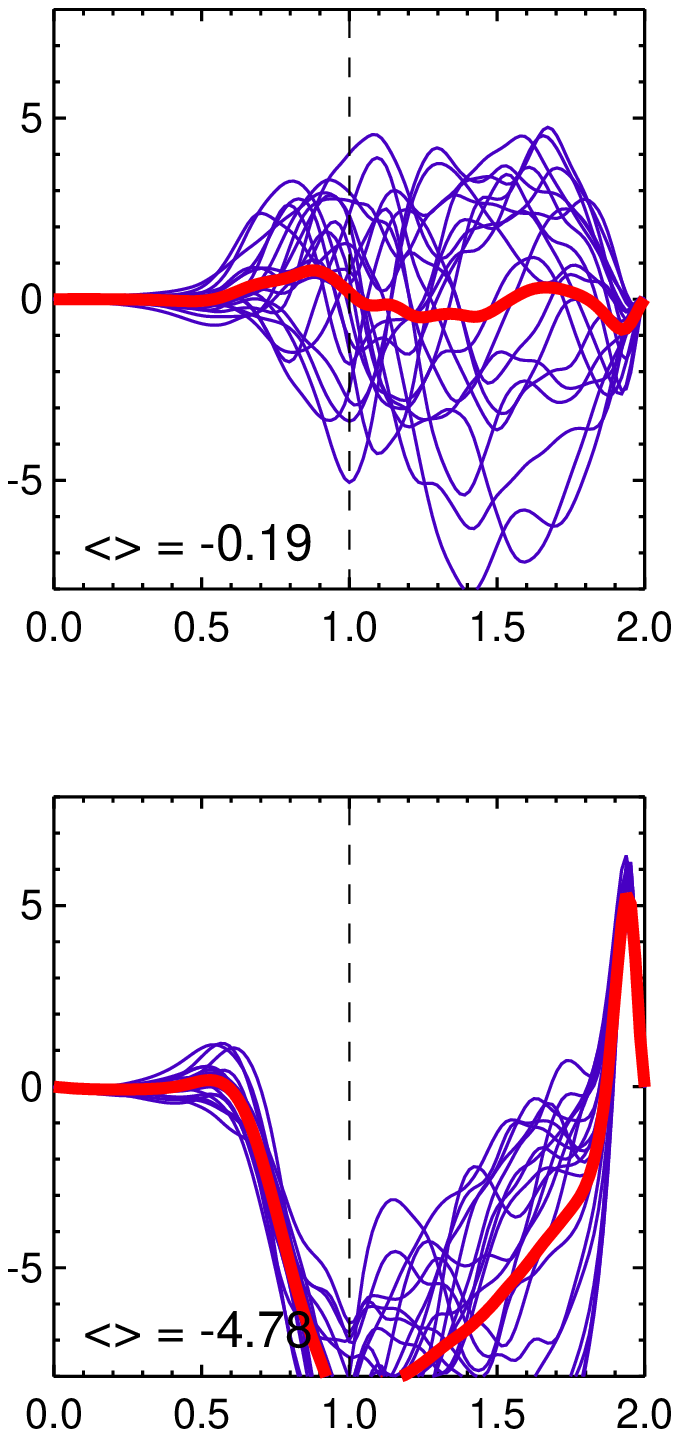}
}
\end{center}
\caption{The two horizontal components (top: ${\cal E}_x$, bottom: ${\cal E}_y$) of the turbulence-induced  EMF for background fields without helicity. Left panel: ${\bar B}_x=0$, $\bar B_y=5$ (only homogeneous magnetic field in $y$ direction); right panel: ${\bar B}_x=10$, $\bar B_y=0$ (only homogeneous electric-current in $y$ direction). The  averages over all snapshots  are noted in the plots.}
\label{fig2}
\end{figure}

Consider first the two different  helicity-free  cases where in the $y$ direction either only a field   or   only  an electric-current exists. Hence, for ${\bar B}_x=0$  
\begin{equation} 
{\cal E}_x\!= \gamma_z {\bar B}_y,\ \ \ \ \ \ \ \ {\cal E}_y=0.
\label{E83} 
\end{equation} 
Figure  \ref{fig2} (left) displays  the resulting $\gamma_z$ as positive (i.e. the flow is downwards) and of order $\gamma_z\simeq 0.24$ in units of  $c_{\rm ac}/100$ which with $c_{\rm ac}\simeq 10$ km/s approximately\footnote{Numerical values in physical units are always given for the top of the unstable zone.}  leads to a downward pumping of   20 m/s. The simulated electromotive force ${\cal E}_y$
 vanishes as required  by (\ref{E83}). 

The alternative example with  $\bar B_y=0$ yields ${\cal E}_x\!=0$ and 
\begin{equation} 
 {\cal E}_y\!=-  \left((\eta_{\rm T} -\hat\eta \frac{\bar B_x^2}{\mu_0\rho})
+\big( \gamma_z-\frac{2}{3} \frac{z}{H^2}  \hat\eta\frac{{\bar B}_x^2}{\mu_0\rho}  \big)z \right)\frac{{\bar B}_x}{H}.
\label{E84} 
\end{equation} 
Indeed, the numerically  simulated  ${\cal E}_x$  vanishes while ${\cal E}_y$ is negative  (see Fig. \ref{fig2}, right). 

\begin{figure}[hbt]
\begin{center}
\hbox{
\includegraphics[width=4.5cm,height=6cm]{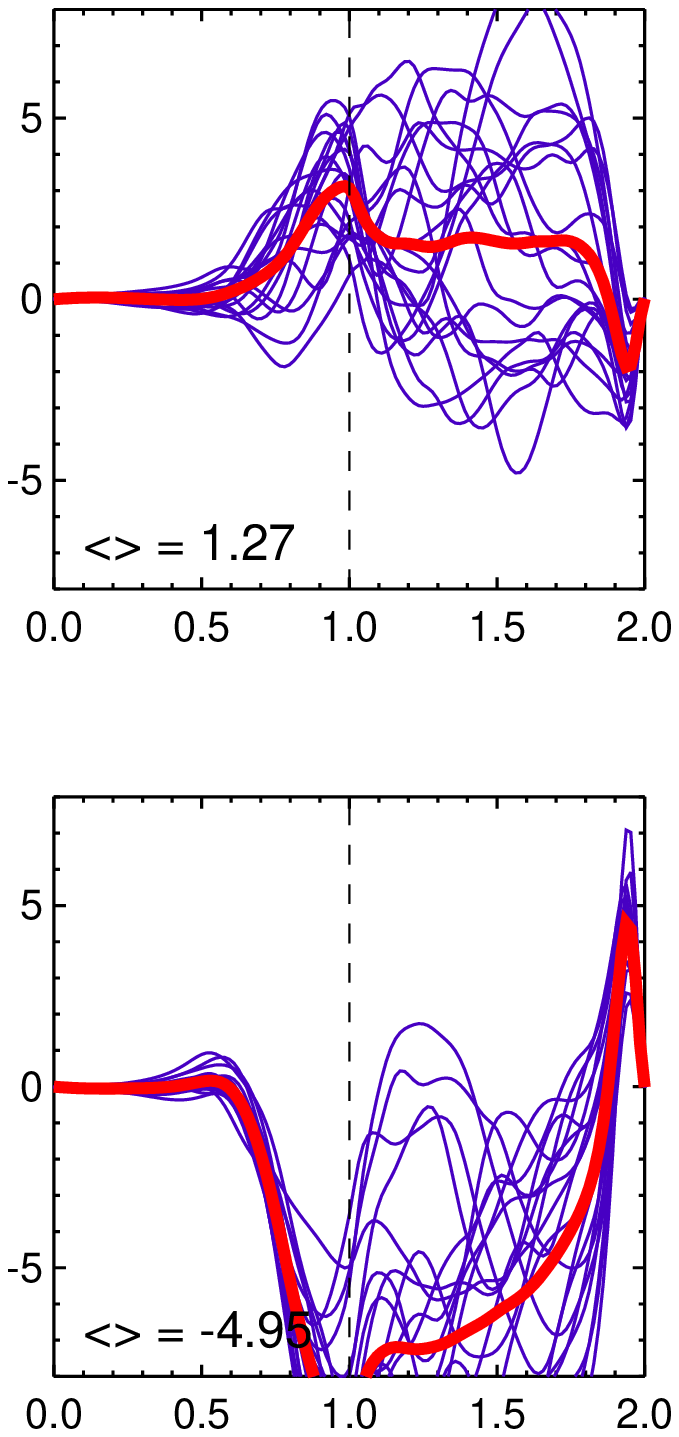}
\includegraphics[width=4.5cm,height=6cm]{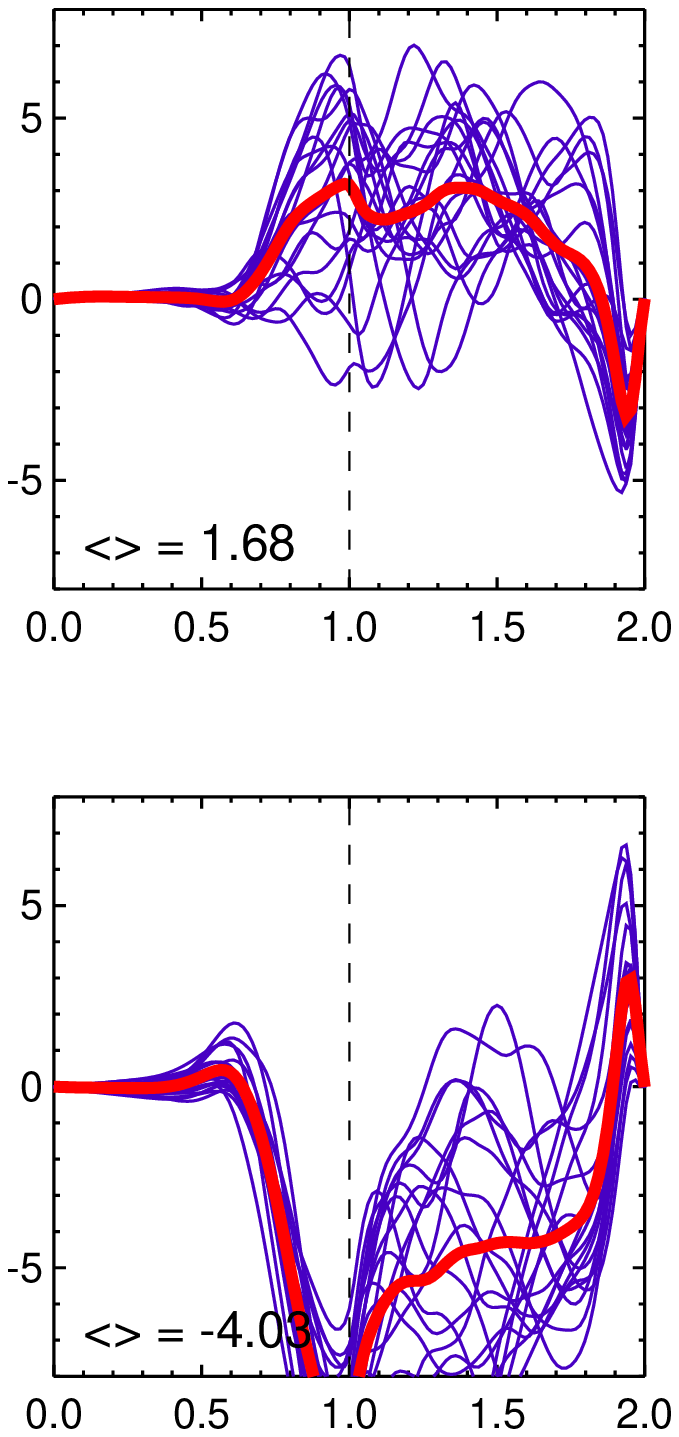}
}
\end{center}
\caption{The same as in Fig. \ref{fig2} but for the helical background fields with ${\bar B}_x=10$, $\bar B_y=5$ (left panel) and   ${\bar B}_x=10$, $\bar B_y=10$ (right panel).  The  averages over all snapshots  are noted in the plots.}
\label{fig3}
\end{figure}

In Fig. \ref{fig3} the two components of the electromotive force are shown for a field with global helicity, i.e. $ {\bar B}_x \bar B_y > 0$.  From (\ref{E4}) one finds
\begin{equation} 
\tilde\gamma_z= \frac{{\cal E}_x}{ {\bar B}_y}, \ \ \ \ \ \ \ \ \ \ \  \ \ \ \ \ \ \ \ \ \ \ \ {\tilde\eta}=-H(\frac{{\cal E}_y}{\bar B_x}
+\frac{{\cal E}_x}{\bar B_y}\frac{z}{H}),
\label{E81} 
\end{equation} 
 where the tildes mark the magnetically quenched values of the pumping velocity and the eddy diffusivity.
The top panels of the plots only concern  the advection velocity $\tilde\gamma_z$ which is always positive. It sinks from $\gamma_z\simeq 0.42 $  to $\tilde \gamma_z\simeq 0.25$  for $B_x=10$ by the magnetic suppression. From the definition of $\tilde \gamma_z$  one immediately finds $\hat\eta\simeq 0.02 $ for the quenching coefficient (\ref{E5})  in units of $100 H/c_{\rm ac}$.  The code works with $\rho_0=1$.

{ Note that in all cases the numerical values of $\langle  {\vec u} \times {\vec b}  \rangle
$ are very small compared with the scalar product $u_{\rm rms}$ and  $b_{\rm rms}$
given in Fig. \ref{fig1}. The resulting correlation coefficients are here very small.

 The existence of an   $\alpha$ effect in the expression  (\ref{E3}) for the electromotive force  would require   a term $ (\bar{\vec{B}} \cdot {\rm curl}\, \bar{\vec{B}})   \bar{\vec{B}}$.   No such term   resulted  from  the quasilinear  approximation for the influence of helical large-scale fields on nonhelically driven turbulence. This finding  can be probed with the numerical simulations.   Note that three models 
   in Figs. \ref{fig2} and \ref{fig3} are computed for constant $B_x$ and for growing helicity  ($ {\bar B}_x \bar B_y=$0, 50, 100).  If an $\alpha$ effect existed one should find that the values of ${\cal E}_y$ grow for growing helicity. The numerical results for the three models, however, are always identical. They do not show  any numerical  response to the growing helicities   indicating the nonexistence of an $\alpha$ term proportional to  $ \bar{\vec{B}} \cdot {\rm curl}~\bar{\vec{B}}$.  The nonexistence  of the $ \alpha$  term  $ (\bar{\vec{B}} \cdot {\rm curl}~\bar{\vec{B}})   \bar{\vec{B}}$ in the electromotive force (\ref{E3}) in the quasilinear theory is thus {  confirmed} by   nonlinear simulations}.

Equation (\ref{E81}) allows to derive the magnetic diffusivity from the simulations.  One finds by elimination of $\hat\eta$  from the definitions the relation $\eta_{\rm T}=\tilde \eta+ (\gamma_z-\tilde \gamma)H$. With the given mean values for all snapshots the  model with $\bar B_y=5$ provides  $\tilde\eta\simeq 0.11$, so that   $\eta_{\rm T}=0.27$ in units of $H c_{\rm ac}/100$   for the 
unquenched   eddy diffusivity. If this result is compared with the traditional estimate $\eta_{\rm T}\simeq (1/3) u_{\rm rms}\ell_{\rm corr}$ one finds $\ell_{\rm corr}/H\simeq 0.1$. On the other hand, if $\tilde\eta=0$ is used as the definition of   a critical magnetic field, i.e.  $B_{\rm eq}=\sqrt{\eta_{\rm T}/\hat\eta}$,  then $B_{\rm eq}\simeq 3.6$ in units of $\sqrt{4\pi\rho} c_{\rm ac}/100$ so that with $\rho\simeq 10^{-2}$ g/cm$^3$ and $c_{\rm ac}=10$ km/s  the  field amplitude  $B_{\rm eq}\simeq 13$ kGauss  results in physical units.
 \subsection{Variation of the magnetic Prandtl number}\label{Prandtl}
It remains to probe the analytical formulation  (\ref{HH}) after which the ratio of the small-scale current helicity and the large-scale helicity grows with the  magnetic Reynolds number. This is certainly only true for large values of $\Rm$ (high-conductivity limit). Large $\Rm$ require high values of the Reynolds number  $\Rey$ and/or not too small $\Pm$. It would certainly be more realistic to increase the Reynolds number rather than the magnetic Prandtl number but our  value $\Rey$ of order $10^3$ is the upper limit defined by technical restrictions. 

Sofar the simulations have been done with a fixed magnetic Prandtl number $\Pm=0.1$. For fixed molecular viscosity its  increase (decrease)  by one order of magnitude effectively increases (decreases) the magnetic Reynolds number by a factor of ten.  We expect a similar  reaction of the small-scale current helicity (\ref{HH}). The reason is that after (\ref{HH}) it grows  for growing  magnetic Reynolds numbers. Because of $\Rm= \Pm\, \Rey$ for fixed  Reynolds number the magnetic Reynolds number runs with $\Pm$. 

\begin{figure}[hbt]
\begin{center}
\hbox{
\includegraphics[width=4.5cm,height=4cm]{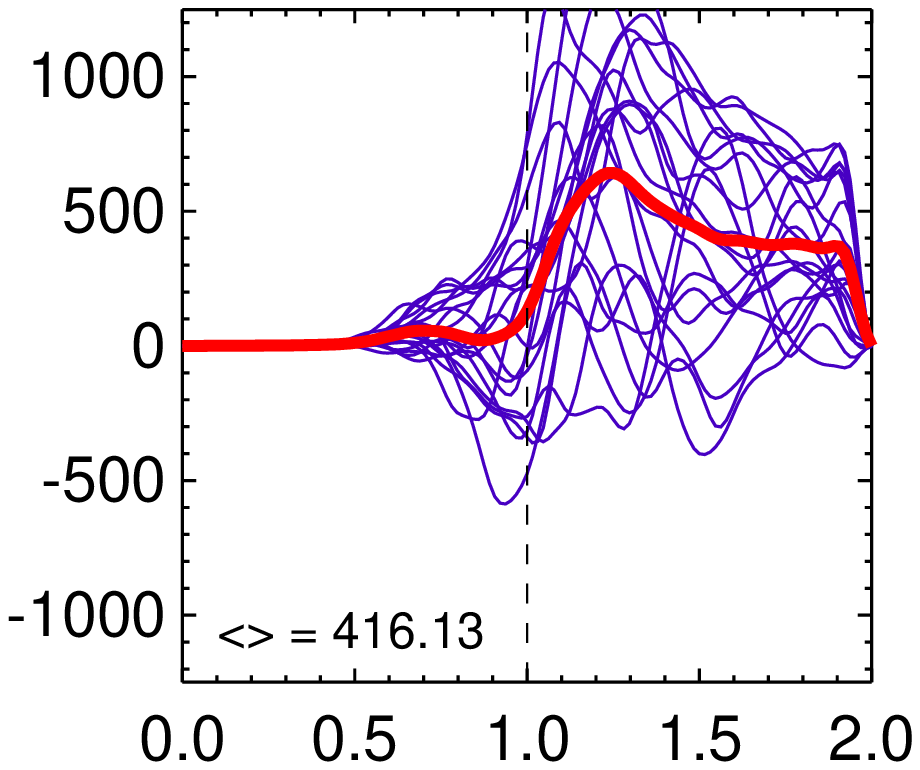}
\includegraphics[width=4.5cm,height=4cm]{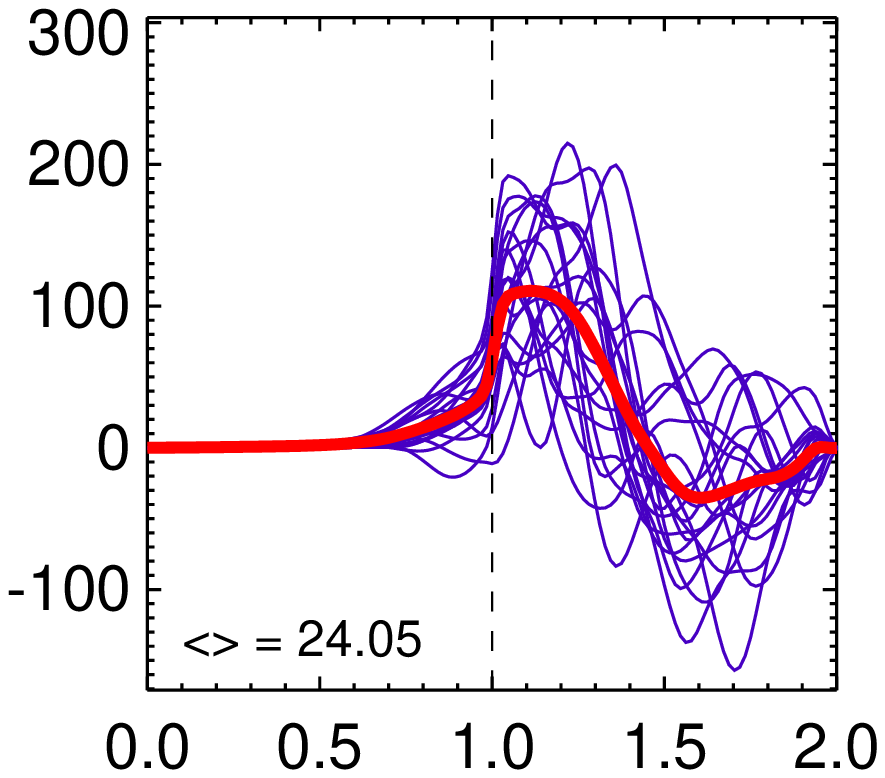}}
\end{center}
\caption{${\cal H}_{\rm curr}$   for   ${\bar B}_x={\bar B}_y=10$.  Left panel:  $\Pm=1$,    right panel:  $\Pm=0.03$.  }
\label{fig5a}
\end{figure}

The results of the variation of the magnetic Prandtl number are shown by the plots given in  Fig. \ref{fig5a}. From left to right the magnetic Reynolds  number sinks by a factor of 33 while the  current helicity sinks by a factor of about  17  (i.e. 50\% of 33). The current helicity  indeed proves to be proportionate to $\Rm$ which, however, proves to be  smaller than the global magnetic Reynolds number of order 100. Since in paper I  for much weaker fields the resulting $\Rm$ were much higher the suggestion may be allowed that for the present  models the magnetic quenching effect is  reducing  the numerical values.

\begin{figure}[hbt]
\begin{center}
\hbox{
\includegraphics[width=4.5cm,height=4cm]{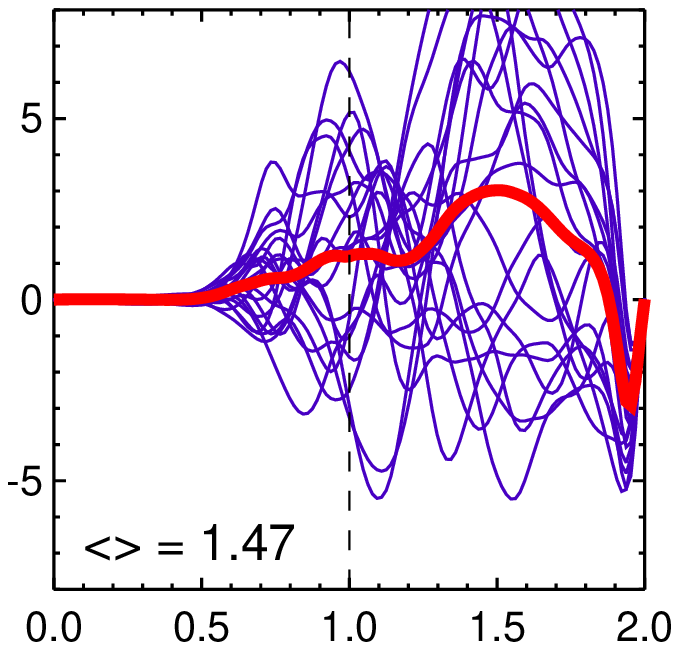}
\includegraphics[width=4.5cm,height=4cm]{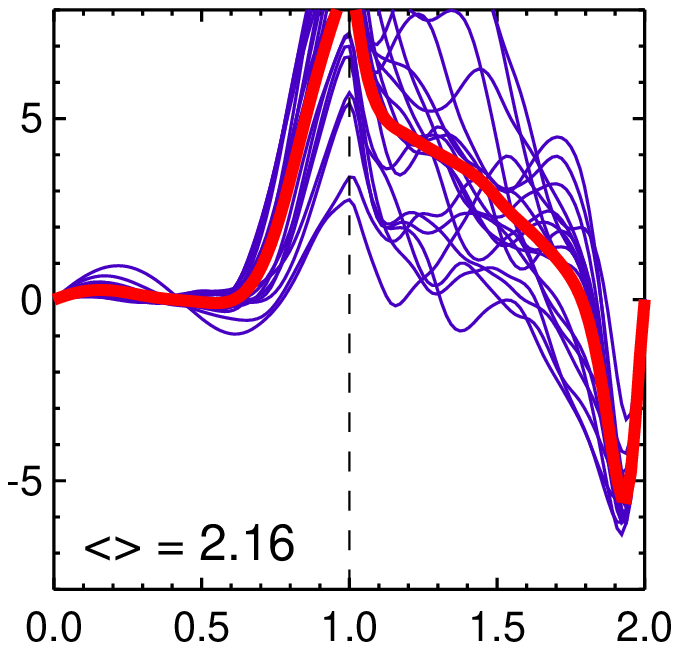}
}
\end{center}
\caption{The same as in Fig. \ref{fig5a} but for ${\cal E}_x$.   }
\label{fig5b}
\end{figure}

It is easy to give   the strength of this statement a test.  It is known that the  advection term ${\gamma_z}$ represented after (\ref{E83}) by  ${\cal E}_x/{\bar B}_y$ does not vary with $\Rm$ (see Kitchatinov et al. 1994).
 Hence, we do {\em not} expect a similar  behavior as in Fig. \ref{fig5a}  for the electromotive force ${\cal E}_x$.
 Figure \ref{fig5b} indeed shows 
 that for small $\Rm$ the reduction of the magnetic Prandtl number by a factor of 33  leaves the electromotive force ${\cal E}_x$  basically unchanged. 
 
 The  results  of Figs. \ref{fig5a} and \ref{fig5b} can also be understood as probing   the inner consistency of  the models and the numerical procedures. It is additionally shown that  basic  results of the quasilinear approximation are confirmed by  simulations. 
 \section{Conclusions}
By analytical (quasilinear) theory and numerical simulations it is  shown that in a nonrotating  convective layer a large-scale helical magnetic background field produces a small-scale current helicity while  the   kinetic helicity vanishes. The current helicity possesses  the same sign as the helicity of the large-scale field.  The ratio of both  pseudoscalars runs with   the magnetic Reynolds number of the convection.  The same magnetic model  provides finite values of the diamagnetic pumping term $\vec{\gamma}$ and the  eddy diffusivity  $\eta_{\rm T}$ but an  $\alpha$ effect does not occur in the same order. 
 
If it is true that the $\alpha$ effect in the stellar convection zones is due to the action of the Coriolis force on the turbulent convection then it is positive (negative)  on the northern (southern) hemisphere which after Eq. (\ref{Kei}) would lead to the opposite signs for the small-scale current  helicity -- as is indeed observed as dominating during the activity  cycle.  The new effect of the small-scale current helicity due to the influence of a  large-scale field  combined with a parallel electric-current     yields    opposite signs.  
During the minimum phase  of the solar activity cycle  just these signs sometimes occur.  Zhang et al.  (2010) give the approximative value $10^{-5}$G$^2$/cm for the observed small-scale current helicity. With the radial scale of 100 Mm and with $\bar{B}_r\bar{B}_\phi\simeq 10^4     $G$^2$ one only needs a magnetic Reynolds number $\Rm$ of the fluctuations of order 10 to fulfill the numerical constraints.

\end{document}